\documentclass[fleqn,usenatbib]{mnras}

\usepackage{newtxtext,newtxmath}

\usepackage[T1]{fontenc}
\usepackage{lscape}

\DeclareRobustCommand{\VAN}[3]{#2}
\let\VANthebibliography\thebibliography
\def\thebibliography{\DeclareRobustCommand{\VAN}[3]{##3}\VANthebibliography}

\usepackage{graphicx}	
\usepackage{amsmath}	
\usepackage{amssymb}	

\usepackage{subcaption}
\usepackage{gensymb}

\title[Meteoroid Close Encounters]{Using Atmospheric Impact Data to Model Meteoroid Close Encounters}

\author[P. M. Shober et al.]{
P. M. Shober,$^{1}$\thanks{E-mail: patrick.shober@postgrad.curtin.edu.au}
T. Jansen-Sturgeon,$^{1}$
P. A. Bland,$^{1}$
H. A. R. Devillepoix,$^{1}$
E. K. Sansom,$^{1}$
\newauthor
M. C. Towner,$^{1}$
M. Cup\'ak,$^{1}$
R. M. Howie,$^{1}$
and B. A. D. Hartig$^{1}$
\\
$^{1}$Space Science \& Technology Centre, School of Earth and Planetary Sciences, Curtin University, GPO Box U1987, Perth, Western Australia 6845, Australia
}

\date{Accepted XXX. Received YYY; in original form ZZZ}

\pubyear{2020}

\begin{document}
\label{firstpage}
\pagerange{\pageref{firstpage}--\pageref{lastpage}}
\maketitle

\begin{abstract}
Based on telescopic observations of Jupiter-family comets (JFCs), there is predicted to be a paucity of objects at sub-kilometre sizes. However, several bright fireballs and some meteorites have been tenuously linked to the JFC population, showing metre-scale objects do exist in this region. In 2017, the Desert Fireball Network (DFN) observed a grazing fireball that redirected a meteoroid from an Apollo-type orbit to a JFC-like orbit. Using orbital data collected by the DFN, in this study, we have generated an artificial dataset of close terrestrial encounters that come within $1.5$~lunar~distances (LD) of the Earth in the size-range of $0.01-100$\,kg. This range of objects is typically too small for telescopic surveys to detect, so using atmospheric impact flux data from fireball observations is currently one of the only ways to characterise these close encounters. Based on this model, we predict that within the considered size-range $2.5\times 10^{8}$ objects ($0.1\%$ of the total flux) from asteroidal orbits ($T_{J}>3$) are annually sent onto JFC-like orbits ($2<T_{J}<3$), with a steady-state population of about $8\times 10^{13}$ objects. Close encounters with the Earth provide another way to transfer material to the JFC region. Additionally, using our model, we found that approximately $1.96\times 10^{7}$\,objects are sent onto Aten-type orbits and $\sim10^{4}$\,objects are ejected from the Solar System annually via a close encounter with the Earth.
\end{abstract}




\section{Introduction}


The diffusion of material out from the main-belt (MB) onto comet-like orbits has been discussed briefly in previous studies \citep{fernandez2002there,fernandez2014assessing,hsieh2016potential,shober2020did}. This mixing potentially can send many durable meteoroids from the MB onto comet-like orbits. Without a clear understanding of this process, meteoroids may be misidentified, leading to inaccurate conclusions about the comet population.

The possibility of already having cometary meteorites in the world's collections has been a topic of discussion for decades \citep{campins1998expected,gounelle2008meteorites}. Jupiter Family Comets (JFCs) are the most likely source region to supply cometary meteorites to the Earth, as the contribution from the nearly isotropic comet (NIC) population is negligible in comparison. However, whether JFCs are capable of producing genetically cometary material on Earth is dependent on the physical lifetimes of JFCs, the dynamic efficiency of JFCs evolving onto Earth-intersecting orbits, and the ability of the meteoroids to eventually survive the atmospheric passage intact as meteorites.

\subsection{Jupiter Family Comets}
Dynamical studies have shown that the scattered disk (SD) and the Kuiper Belt are the two primary sources for modern JFCs  \citep{levison1997kuiper,duncan1997disk}. The SD is the most dominant as the Kuiper Belt is slower at producing JFCs. The SD is cold, likely very primitive and volatile-rich, with larger eccentricities than the classical Kuiper Belt \citep{gomes2008scattered}. Other sources from within the MB have also been proposed to partially supply material to the JFCs \citep{fernandez2002there,kim2014physical,fernandez2015jupiter,hsieh2020potential}. This can occur as a result of outward diffusion from the MB via mean-motion resonances (MMR), primarily the 2:1\,MMR ($3.27$\,au). Other outer-MB resonances such as the 9:4 and 11:5 MMRs have been suggested to be able to produce a modest amount of objects on JFC-like orbits \citep{fernandez2014assessing,fernandez2015jupiter,hsieh2020potential}. Studies have additionally found that terrestrial planets (particularly Earth and Venus) may play an important role by perturbing MB objects onto JFC-like orbits \citep{fernandez2002there,hsieh2020potential}.


Two primary factors are responsible for the observed JFC size distribution: the size distribution for the source regions of the JFCs and the physical evolution that bodies on JFC orbits underwent. Several studies have attempted to characterize the JFC cumulative size-frequency distribution (CSD) \citep{meech2004comet,fernandez2013thermal}. The CSDs determined in these studies, based on telescopic observations, predict a break in the slope for sub-kilometre bodies. This break is due to the obvious sampling bias against telescopically observing these objects. However, \citet{meech2004comet} argued that despite this sampling bias, there should still be more discovered sub-kilometre JFC objects according to their model. The most likely explanation for this paucity of sub-kilometre objects is the short physical lifetimes associated with these objects. While the average dynamical lifetime for JFCs is typically $10^{5}$~years, there is ample evidence that the physical lifetimes are $\sim10^{3}$~years for JFCs in the inner solar system \citep{kresak1981lifetimes,kresak1990secular,levison1997kuiper,fernandez1999population,hughes2003variation,disisto2009population,sosa2012asymmetric}. 

\subsection{Meteors and Fireballs}
On the other end of the spectrum, ground-based meteor and fireball observation networks are able to characterise the smallest subset of objects on JFC-like orbits. There are several meteor showers observed to originate from JFCs (e.g., Draconids and Andromedids). These showers tend have larger meteoroids than long-period comets (LPCs) ($\sim100\,\mu$m) and lower impact speeds ($11-35$\,km\,$s^{-1}$) \citep{jenniskens2006meteor}. In the study of the zodiacal cloud by \citet{nesvorny2010cometary}, they concluded that particles from JFCs should represent $85\%$ of the total terrestrial mass influx. This result provides an explanation for the abundance of micrometeorites with primitive carbonaceous compositions found in Antarctica. Though, as the meteoroid sizes increase to centimetre and metre scales, the story becomes uncertain. 

Despite the predicted paucity of sub-kilometre objects based on CSDs of the JFC population, there have been many bright fireballs observed to originate from JFC-like orbits. These meteoroids can be centimetres to metres in scale. However, the delivery mechanism and composition of these objects is still unclear. For example, \citet{madiedo2014b} reported observing a bright fireball originating from a JFC orbit with mass of $40\pm5$\,kg. The meteoroid penetrated as deep as $68$\,km altitude and had a maximum luminosity of $-13\pm0.5$ absolute magnitude. However, the object could not be associated with any known JFCs. If the object was genetically JFC material ($T_{J}=2.3\pm0.2$), given its low perihelion distance, the meteoroid is predicted to have an extremely short physical lifetime. The physical lifetimes for kilometre-size JFCs is typically on the order of $10^{3}$\,years, however, metre-sized fragments are estimated to only persist for a few revolutions ($\leq10$\,yrs) \citep{beech2001endurance}. Therefore, either it did not originate from the Jupiter family comet population or there is a mechanism capable of extending the physical lifetimes of JFC material in this size-range. \citet{brown2016orbital} also analysed 59 fireballs caused by meteoroids $\geq1$\,m in diameter, and found $10-15\%$ have a possibly cometary origin, but only about half of these were observed to be weaker than average based on ablation behavior. Additionally, in \citet{flynn2018physical}, they compared the connection between $T_{J}$ and PE criterion for 600 fireball observations showing that fireballs with JFC-like $T_{J}$ values display a similar spectrum of PE values to those of meteoroids from asteroidal orbits ($T_{J} > 3$). The catastrophic breakup or splitting of the parent body is currently the most favoured explanation to produce large fragments from a comet, as no other mechanism is capable of producing debris in this size-range \citep{jenniskens2004,jenniskens2005meteor}.

\subsection{Meteorite Falls}
There have been a handful of meteorite falls associated with JFC-like orbits. Nearly all of these have likely origins in the MB, given they all have $T_{J}\sim3$. \citet{granvik2018identification} re-calculated the source region probabilities for 25 meteorite falls and found only one meteorite, Ejby, to have its most likely source be the JFCs. Curiously, Ejby and Ko\u{s}ice, the two meteorites with the highest likelihoods of originating from the JFC source region, are both H-chondrites. Meanwhile, two CM-chondrites (Maribo and Sutter's Mill) both have non-negligible chances of coming from the JFC population, but are most likely sourced from the MB via the $3:1$ resonance. 

\subsection{Addressing the Problem}
Telescopic observations of JFCs display a paucity at sub-kilometre scales. When observing dust-sized objects, primitive CM-chondritic material is abundant and associated with JFCs \citep{nesvorny2010cometary}. Nevertheless, fireball networks report metre-scale objects on JFC orbits impacting the Earth. If asteroidal material from the MB is regularly transferred to the JFC population, given the extremely short physical lifetimes expected for cometary meteoroids, the fireballs from non-shower JFC-like orbits may be highly contaminated by asteroidal material. 

Objects that are $>100$\,m in diameter regularly come close to the Earth\footnote{https://cneos.jpl.nasa.gov/ca/}. However, these bodies are also much easier to identify telescopically. Thus, there is certainly a much larger group of smaller objects that are not seen by telescopes that pass very close to the Earth. This smaller-size subset encompasses the size-range of objects that typically generate meteorites. In a previous study, \citet{shober2020did} described a grazing fireball event that transferred a meteoroid from an Apollo-type orbit to a JFC-like orbit. Since there are many objects like this that go unnoticed by telescope surveys, there is likely a non-negligible amount of small objects that are quickly inserted into dynamically distinct orbits. Additionally, objects tend to evolve along the lines of equal perihelion or aphelion. Therefore, the probability of re-observing these gravitationally scattered objects may be worth consideration. Close encounters with the Earth could provide an additional way to transfer material from the MB to JFC-like orbits \citep{fernandez2015jupiter,hsieh2020potential}.\\

\textbf{Objectives for this study:}
\begin{itemize}
    \item Simulate the close encounter population with the Earth based on the DFN dataset. 
    \item Identify how the orbits have changed as a result of the close encounters with the Earth.
    \item Characterise the sub-population of objects redirected onto JFC-like orbits ($2<T_{J}<3$) from asteroid-like orbits ($T_{J}>3$). 
    \item Estimate the impact frequency of objects redirected from asteroid-like ($T_{J}>3$) to JFC-like orbits ($2<T_{J}<3$); i.e., determine whether this population impacts the Earth frequently enough to be observed by fireball networks. 
\end{itemize}

\section{Methods}

\subsection{Desert Fireball Network Data}
The DFN is part of the Global Fireball Observatory (GFO), a multi-institutional collaboration of partner fireball networks around the Earth. The DFN, the largest single fireball network in the world, covers about one-third of Australian skies every night using automated high-resolution digital fireball observatories \citep{bland2012australian,howie2017build}. The DFN collects massive amounts of all-sky imagery that is automatically processed, producing a highly accurate orbital dataset of fireballs \citep{howie2017submillisecond,sansom2015novel,jansen2019comparing,towner2019fireball,sansom20193d,sansom2019determining}. 

\begin{figure}
    \centering
    \includegraphics[width=\columnwidth]{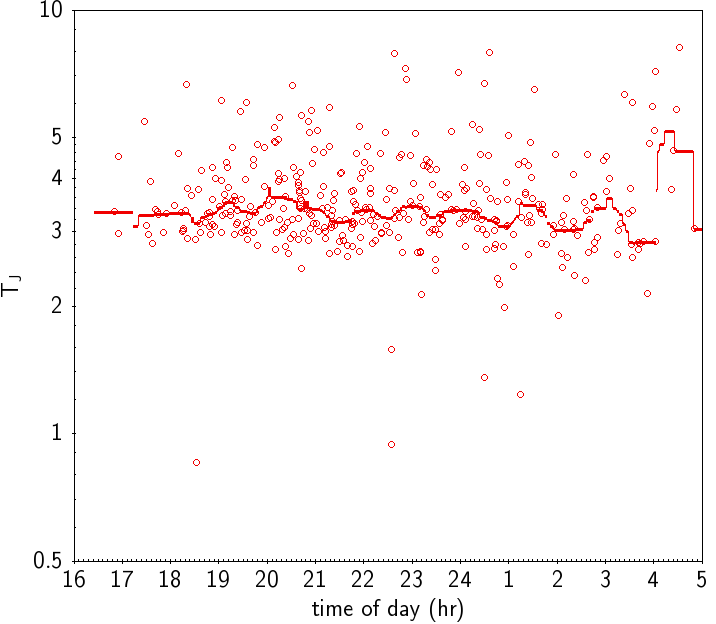}
    \caption{Tisserand`s parameters ($T_{J}$) of the DFN events used in this study along with the time of day in which each fireball was observed. The red line represents the median $T_{J}$ value. }
    \label{fig:24hr_tj}
\end{figure}

\begin{figure}
    \centering
    \includegraphics[width=\columnwidth]{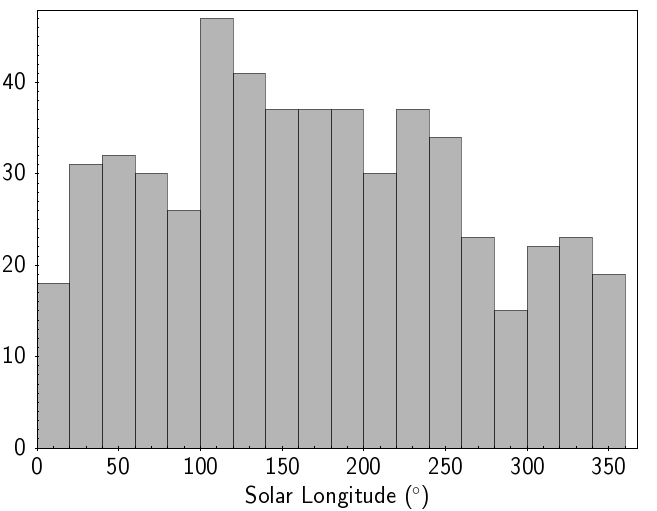}
    \caption{Solar longitude distribution for DFN events used to generate artificial close encounter population. There is a noticeable decrease in events detected during the summer months as the duration of the night is shortest.}
    \label{fig:solar_long}
\end{figure}

\subsection{Addressing Observational Biases}
In this study, we will be producing an artificial dataset of close encounters with the Earth of centimetre to metre-sized objects using data collected by the DFN. The flux of objects observed by the DFN for the previous four years was employed to construct a model of the close encounter population. However, in order to adequately estimate this population, we must first address the intrinsic observational biases of the DFN. Biases are listed below along with how each were considered in this study: 
\begin{enumerate}
    \item Observations are optimised for the size-range specific to meteorite dropping events.
    \begin{itemize}
        \item The DFN was designed to observe meteorite-dropping fireball events (fireball limiting magnitude $\sim0.5$) \citep{howie2017build}. Thus our dataset represents a subset of larger objects compared to other meteor networks. However, in this study, we are interested in the close encounters of centimetre to metre-sized objects, so this bias is what makes the DFN dataset a good way of understanding these encounters.
    \end{itemize}
    \item The DFN exclusively observes at night; i.e. the antihelion direction. 
    \begin{itemize}
        \item We anticipate that this bias will have a negligible effect on the observed population of objects, as it is primarily due to small changes in the orbital geometry at impact \citep{halliday1982study}. As seen in Fig.~\ref{fig:24hr_tj}, the median Tisserand`s parameter does not vary significantly over the course of the night. There is a single aberration, but this is due to low-statistics in this subset only around 4\,am. 
    \end{itemize}
    \item The DFN fireball dataset contains some meteor showers in addition to sporadic events. 
    \begin{itemize}
        \item The methodology used explicitly assumes there are meteoroids in similar nearby orbits. This assumption for the sporadic meteoroids may be an oversimplification. However, given we have the largest self-consistent fireball dataset in the world (most of which are sporadic), this should reduce the bias as much as possible.  
    \end{itemize}
    \item The DFN observations vary annually with the seasons.
    \begin{itemize}
        \item  Seasonal weather variation can also have a significant effect on observations, notably cloud cover. However, given the continental-scale and dry climate of Australia, the DFN's four years of observations are negligibly affected. As seen in Fig.~\ref{fig:solar_long}, even though the weather is more inclement during the winter months, there is no effect on the observations. There are, in fact, more observations during the winter, due to the longer nights. The amount of observable hours due to changes in daytime length varies about $\pm25\%$ for DFN stations annually. This significantly decreases the total number of events observed during the summer; however, $\sim91\%$ of the events used in this study are sporadic. Thus, since the vast majority of the events are not affected by seasonal biases, we have decided to ignore it. This will cause only a minor underestimate in the summer shower contribution to the close encounter flux. 
    \end{itemize}
    \item The sensitivity of the DFN cameras varies due to whether the Moon is above or below the horizon. 
    \begin{itemize}
        \item The DFN cameras have lower sensitivity when the Moon is above the horizon and higher sensitivity when it is below. The minimum-size cutoff in this study ($0.01$\,kg) should eliminate these monthly variations as we typically observe masses down to $\sim1\text{-}2$\,g on moonlit nights. However, there is expected to be a slight underestimate for objects that are tens of grams in mass.
    \end{itemize}
    \item Observed meteor velocities vary nightly due to changes in the viewing orientation relative to the Earth's motion around the Sun. 
    \begin{itemize}
        \item This bias should not affect observations on an annual level.
    \end{itemize}
    \item The lower end of the observed size-range varies with impact speed. 
    \begin{itemize}
        \item As noted by \citet{vida_et_al_2018}, the minimum size object observed by meteor networks decreases as the velocity increases. Nevertheless, this observational bias can be mitigated by setting a minimum size limit when building the artificial dataset in this study ($0.01$\,kg). This mass limit was chosen as it is the limiting mass observed by the DFN. Events in the dataset with nominal masses less than this are likely inaccurate. 
    \end{itemize}
    \item Due to gravitational focusing, the Earth is hit by a higher proportion of slower objects. 
    \begin{itemize}
        \item When objects approach the Earth at slower relative velocities, they tend to be more gravitationally focused towards the Earth; slightly increasing the ratio of slow impactors observed. Neglecting to account for gravitational focusing would lead to a slight overestimate of the proportion of slow close encounters. In this study, we used the formulation for the enhancement factor by \citet{opik1951} to account for the gravitational focusing. The enhancement factor is thus defined as: 
        
        \begin{equation} \label{eq:1} H_{F}=\frac{A_{G}}{A_{NG}}=\left(1 + \frac{V^{2}_{esc}}{V^{2}_{i}}\right)
        \end{equation}
        
        where $A_{G}$ is the effective cross-sectional area of the Earth due to gravitational focusing, and $A_{NG}$ is the physical cross-section of the Earth. This ratio can be estimated using the escape velocity at the surface of the Earth ($V_{esc}$) and the initial velocity of the meteoroid before the encounter ($V_{i}$). This enhancement factor was used to normalise the artificial close encounter flux generated, producing better relative abundances of certain kinds of encounters (i.e., not overestimating the number of slow-approachers).
    \end{itemize}
\end{enumerate}

\subsection{Creating the Close Encounter Dataset}
The DFN is designed to observe and triangulate fireballs over Australia. Using the data collected over the last four years, assuming that the flux is reasonably representative of the global flux, we can extrapolate outward and try to characterise the close encounter flux for this size range. The events used to generate our model were limited to those with a predicted pre-atmospheric mass ranging from $0.01-100$\,kg. This size range was chosen to limit bias at small sizes, and it is the most well-measured range within the DFN dataset. This results in a set of 581 fireball events with which to initiate our model. Of these events, 50 ($\sim9\%$) are associated with established and non-established showers.

For each event, the predicted state of the meteoroid at the beginning of the luminous phase was integrated back until it was at least three lunar distances (LD) away from the Earth. This procedure was chosen in order to primarily limit the close encounters to a minimum orbital intersection distance (MOID) of within about 1.5\,LD, as the orbits will be decreasingly affected with larger MOID values. For the orbital integrations, we used the IAS15 integrator described in \citet{2015MNRAS_rein} including perturbations from all the planets in the Solar System as well as the Moon. At this point, a cloud of 4000\,particles were generated uniformly by varying the position relative to the actual prediction of the meteoroid by $\pm1.0$\,LD in each Cartesian direction in the heliocentric frame, without varying the velocity. A Hill radius for the Earth is $\sim3.89$\,LD, however we decided to integrate slightly less to save on computation time. We justified this decision by finding that, for all DFN events, the Tisserand's parameter is typically $>99\%$ similar to its pre-encounter value after reaching two LD away from the Earth.

These particles were then integrated forward in time until they were at least three LD away from the Earth again. Any particles which came within 200\,km of the Earth's surface were removed from the simulation, effectively removing impacts and grazing events. In total, 2.3 million particles were integrated to estimate the close encounter flux. This flux was then normalised using the estimated enhancement factor for each event. 

In this model, we explicitly assume that each particle in the simulation represents a group of meteoroids. This is done to make the simulation feasible, as there are expected to be hundreds of billions of close encounters within this size-range annually. The size of the base dataset used (581 fireballs) is very small compared to the number of objects having close encounters with the Earth, however, the general trends observed by the model are likely representative. 

In order to estimate the close encounter flux of the Earth, we derived the following equation to calculate the cumulative flux distribution: 
\begin{equation} \label{eq:2} 
Flux(r)=\left(\frac{r^{3}}{r_{\earth}^{3}}\right) \times N_{H_{F}} \times \iota_{T}
\end{equation}
\begin{equation} \label{eq:3} 
where \quad N_{H_{F}} =\left[1+(H_{F}-1)\left(\frac{r_{\earth}^{3}}{r^{3}}\right)\right]
\end{equation}
$N_{H_{F}}$ is the scaling factor included to account for the effect of gravitational enhancement at smaller geocentric distances, $r$ is the distance from the Earth's centre, $r_{\earth}$ is the radius of the Earth, $H_{F}$ is the enhancement factor, and $\iota_{T}$ is the total terrestrial flux estimated for the top of the atmosphere by \citet{bland2006rate}. As $r \to r_{\earth}$ the scaling factor $N_{H_{F}} \to H_{F}$, whereas as $r \to \infty$ then $N_{H_{F}} \to 1.0$ (i.e., no enhancement when encounters are more distant).

We then calculated the cumulative flux for each particle's MOID, and took the difference between each cumulative flux value to determine the number of objects represented by each particle. The mass and mass errors were determined using the pre-atmospheric masses and errors calculated by the DFN in combination with the terrestrial mass flux estimate for the given size range ($0.01-100$\,kg) \citep{sansom2015novel,bland2006rate}. 

\section{Results and Discussion}

\begin{figure}
     \centering
     \begin{subfigure}[b]{\columnwidth}
         \centering
         \includegraphics[width=\columnwidth]{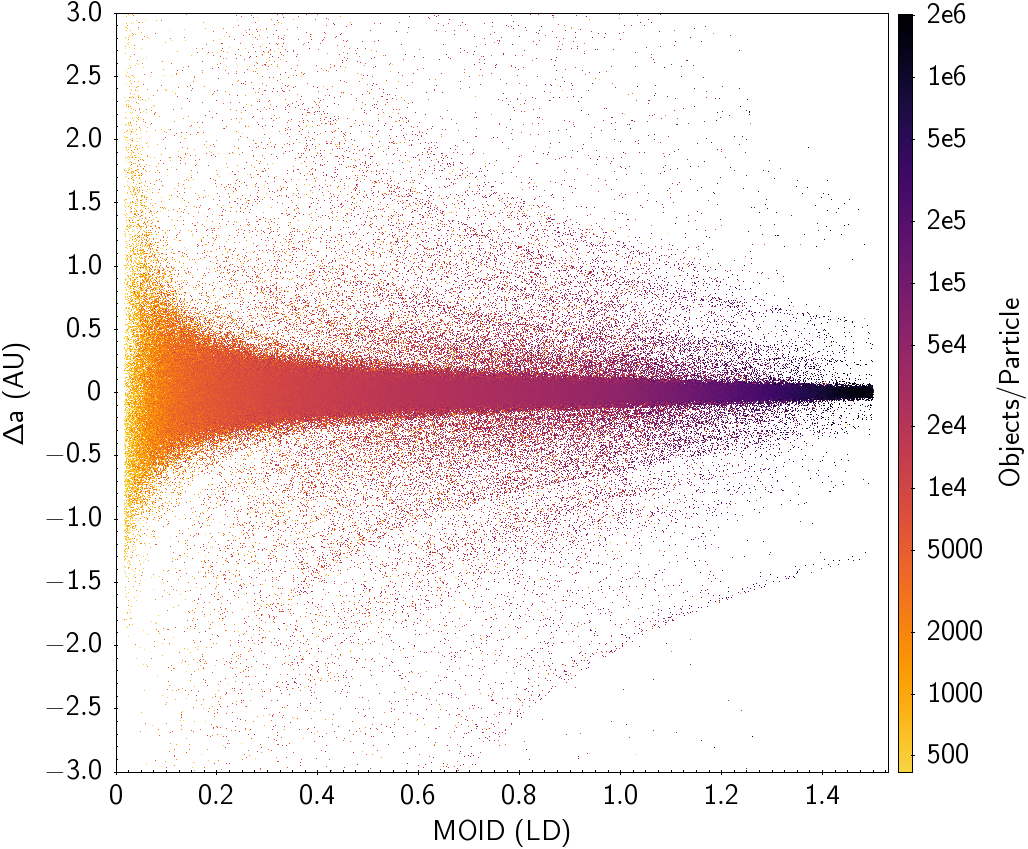}
         \label{fig:a_moid}
     \end{subfigure}
     \hfill
     \begin{subfigure}[b]{\columnwidth}
         \centering
         \includegraphics[width=\columnwidth]{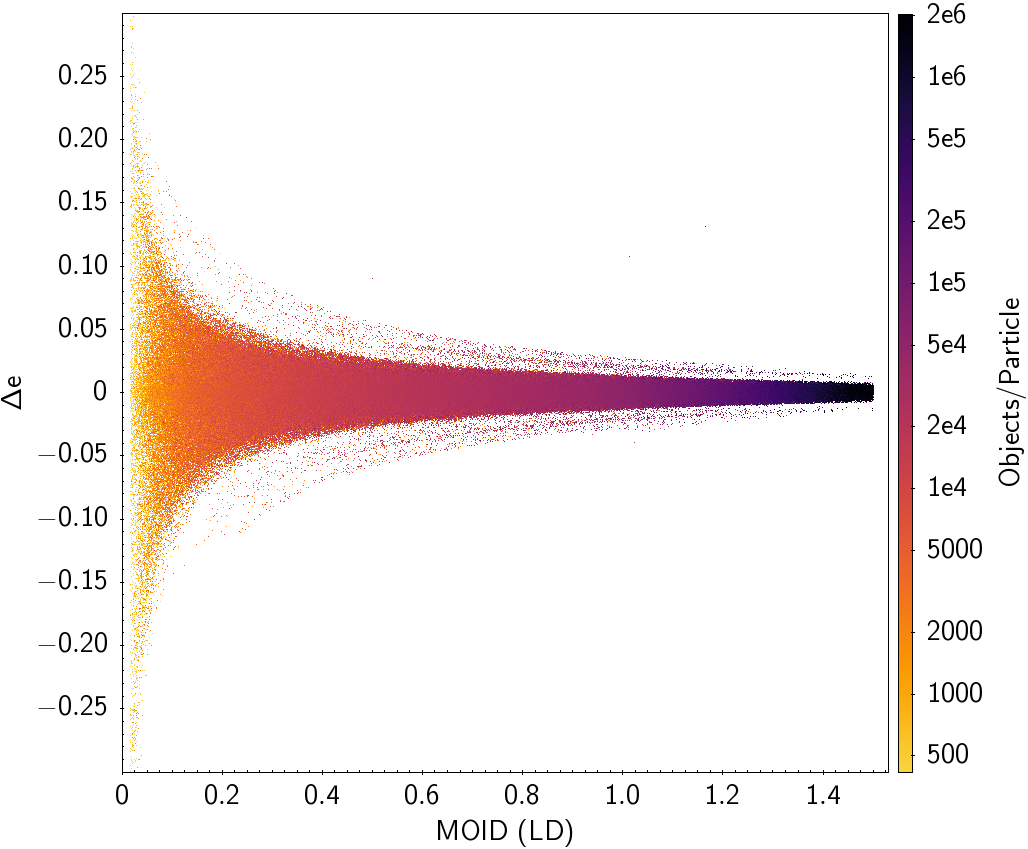}
         \label{fig:e_moid}
     \end{subfigure}
     \hfill
     \begin{subfigure}[b]{\columnwidth}
         \centering
         \includegraphics[width=\columnwidth]{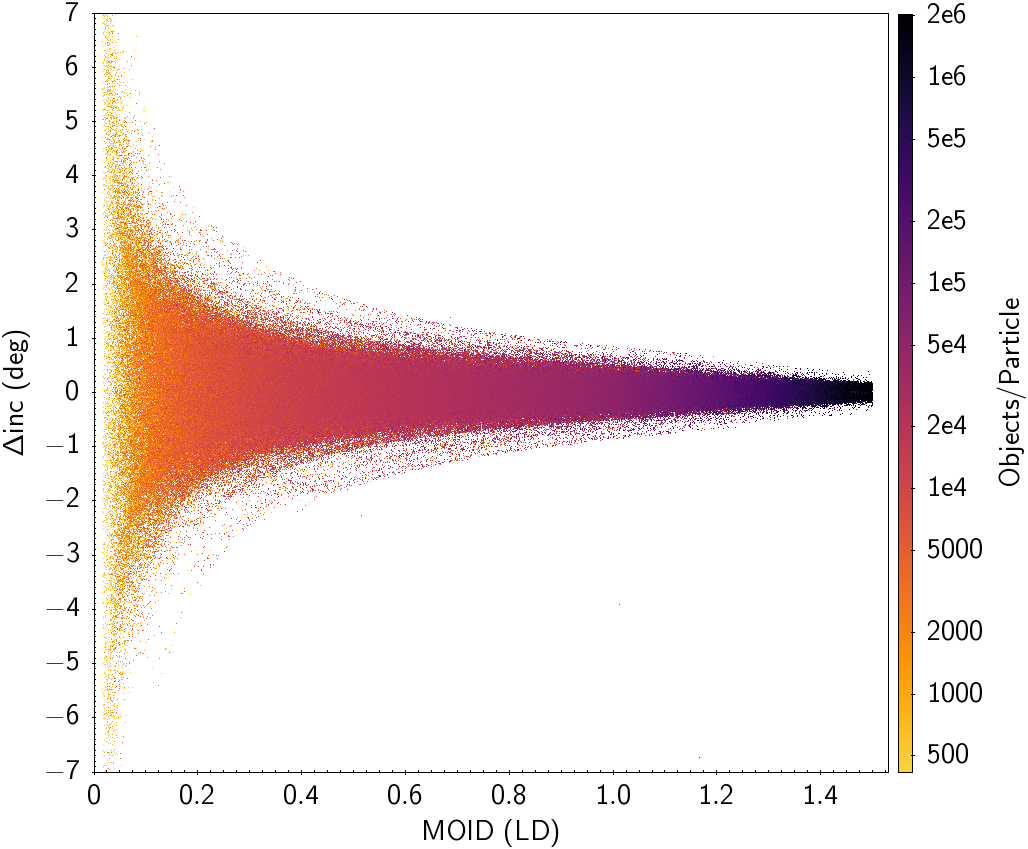}
         \label{fig:inc_moid}
     \end{subfigure}
        \caption{Change in semi-major axis (a), eccentricity (e), and inclination (inc) for every particle in our simulation as a function of the minimum orbital intersection distance (MOID) in Lunar Distances (LD). The color bar indicates the number of objects each particle represents annually.}
        \label{fig:aei_moid}
\end{figure}

\subsection{Orbital Changes} 
Using global flux estimates from \citet{bland2006rate} along with DFN data to extrapolate, we were able to characterise the $\sim10^{11}$ predicted close encounters that occur annually within $1.5$\,LD ranging from $0.01-100$\,kg. A vast majority of the objects that encounter the Earth do not have appreciable changes to their orbits. As seen in Fig.~\ref{fig:aei_moid}, the change in orbital elements are centred on zero with greater magnitude alterations occurring at smaller MOID values ($1/r^{2}$ relationship). Every object gains or loses energy, but usually a negligible amount. For example, for all objects coming within $1.5$\,LD, the median changes found in the semi-major axis, eccentricity, and inclination were $0.019$\,AU, $0.0022$, and $0.11\degree$ respectively. However, for encounters within $0.1$\,LD, the median changes were $0.27$\,AU, $0.033$, and $1.6\degree$ respectively.

\begin{figure}
    \centering
    \includegraphics[width=\columnwidth]{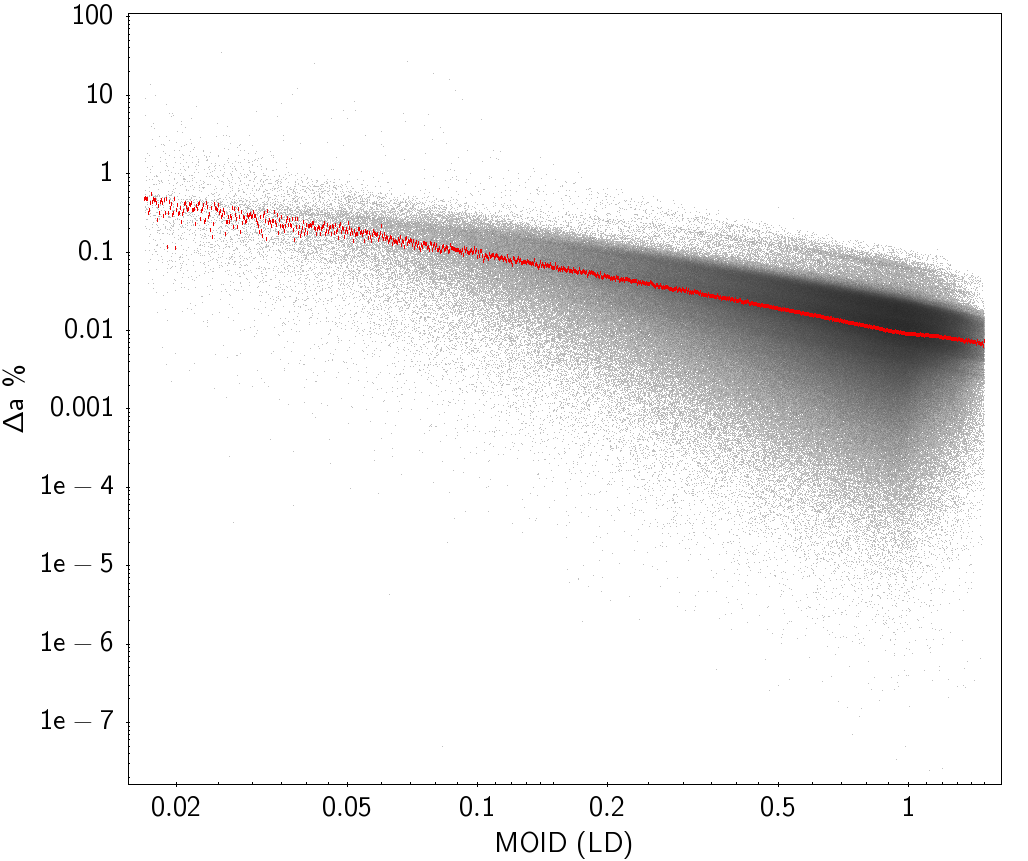}
    \caption{Percentage change in the particles' semi-major axis (a) varying according to the minimum orbital intersection distance (MOID). The red points indicate the median change. The largest change in orbital energy is logically experienced by objects with the closest encounters.}
    \label{fig:moid_da}
\end{figure}

\begin{figure}
     \centering
     \begin{subfigure}[b]{\columnwidth}
         \centering
         \includegraphics[width=\columnwidth]{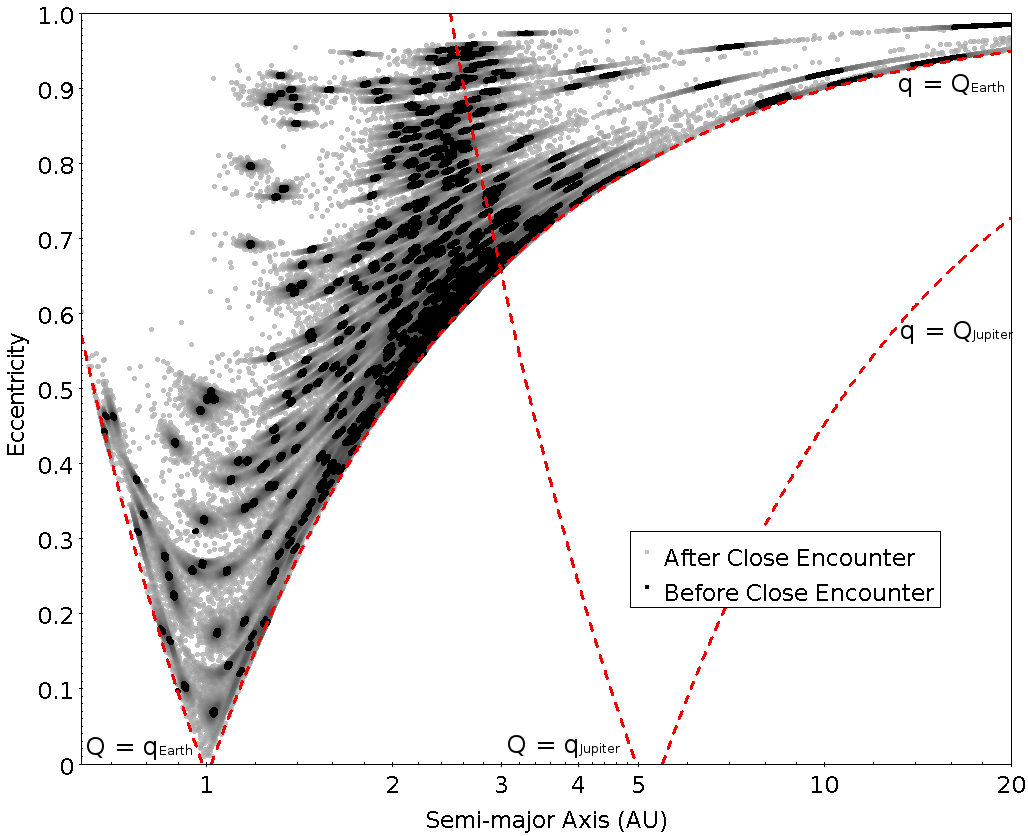}
         \caption{}
         \label{fig:ae_ba}
     \end{subfigure}
     \hfill
     \begin{subfigure}[b]{\columnwidth}
         \centering
         \includegraphics[width=\columnwidth]{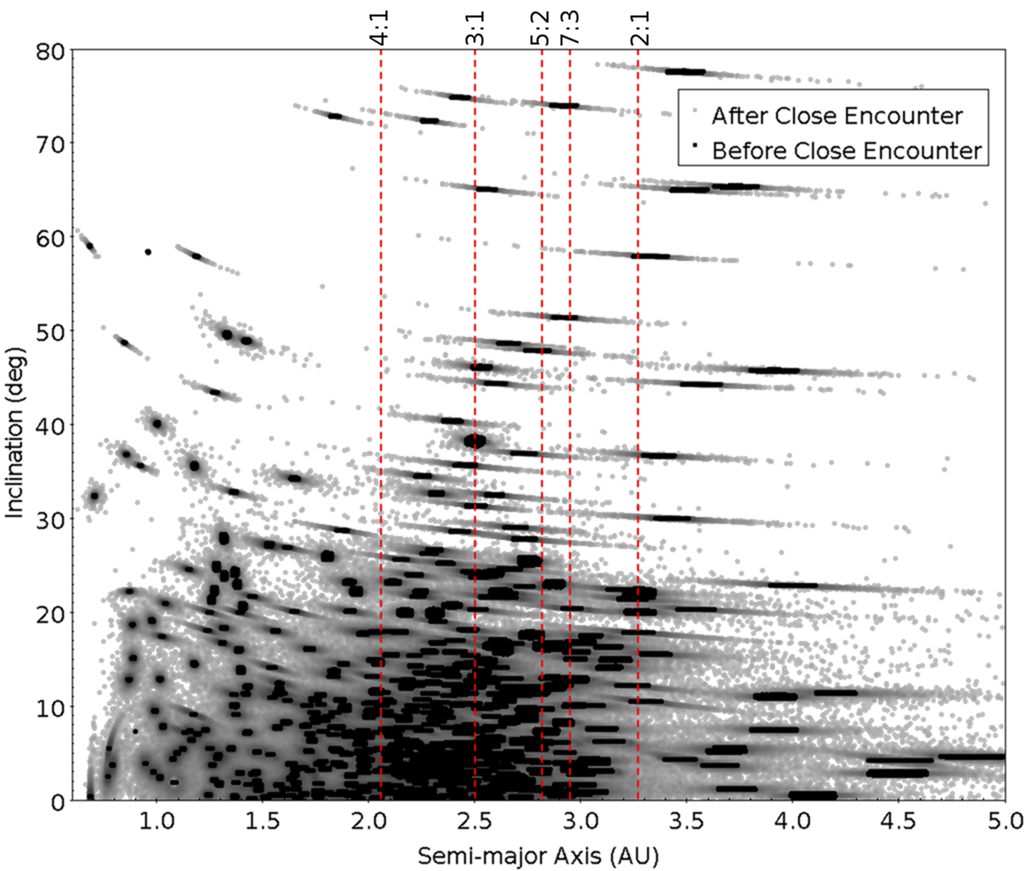}
         \caption{}
         \label{fig:ai_ba}
     \end{subfigure}
        \caption{\ref{fig:ae_ba} and \ref{fig:ai_ba} show the semi-major axis vs. eccentricity and inclination respectively for all the test particles in the model (2.3 million particles) before and after having a close encounter with the Earth. The grouping of black particles tend to get dispersed along lines of equal aphelion/perihelion after having a close encounter (grey particles). There is typically minimal change in the inclination, however, as $a\sim1$ the changes in inclination becomes more significant.}
        \label{fig:aei_ba}
\end{figure}

\begin{figure}
    \centering
    \includegraphics[width=\columnwidth]{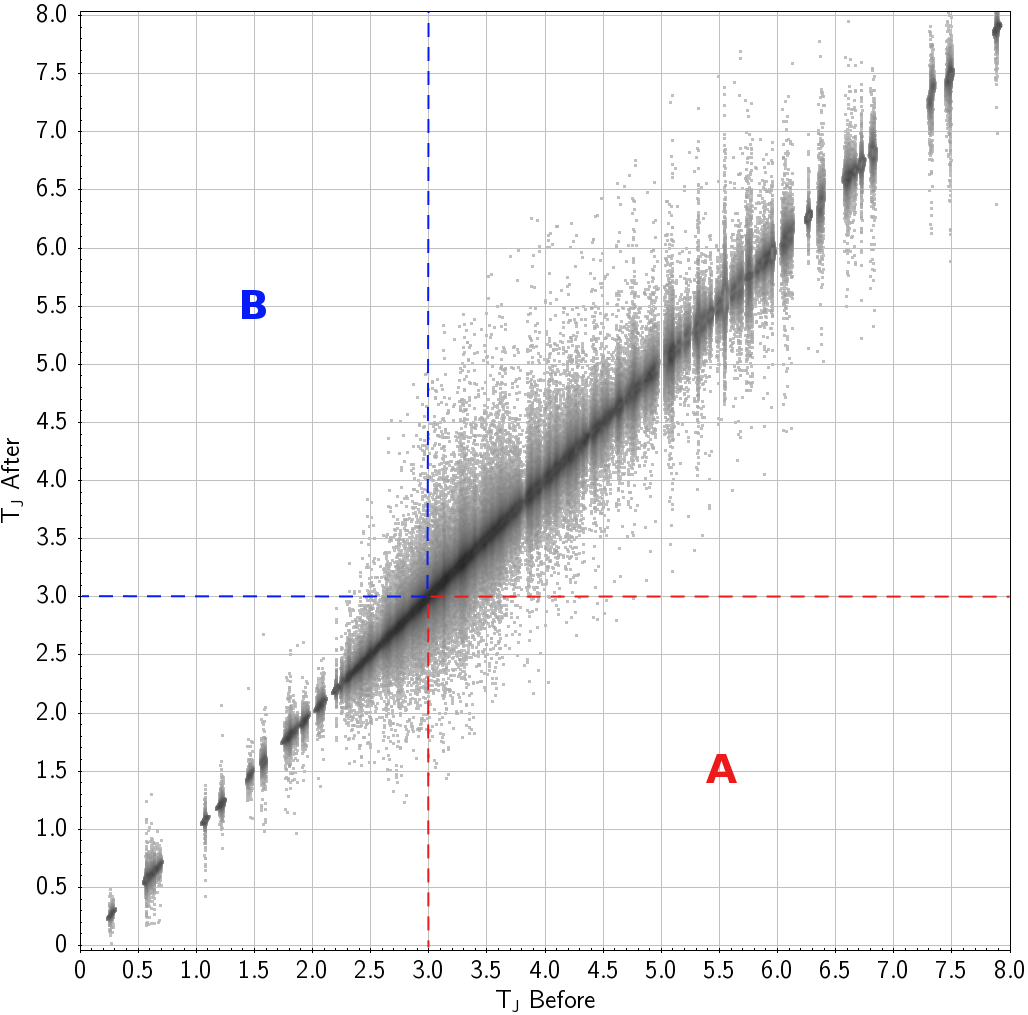}
    \caption{Close encounter simulation particles energy change in terms of Tisserand's parameter ($T_{J}$). The $T_{J}$ value before and after encountering the Earth are the x and y axis respectively. Region A denotes contains particles that were transferred from a asteroid-like orbit to a JFC-like one, whereas region B contains particles that underwent the reverse transfer.}
    \label{fig:tj_ba_scatter}
\end{figure}

The two factors controlling this process are the MOID of the encounter and the pre-encounter velocity of the meteoroid, as seen in Fig.~\ref{fig:moid_da}. This change in the orbital parameters, as shown in Fig.~\ref{fig:aei_ba}, diffuses the objects along the lines of equal perihelion or aphelion. As the black `clouds' of particles generated in the model encounter the Earth, they diffuse outward (grey) where they could be inserted into a resonance or the path of another planet (Fig.~\ref{fig:ai_ba}). If these alterations are significant, the meteoroid's $T_{J}$ value could change enough to dynamically re-classify the orbit (Fig.~\ref{fig:tj_ba_scatter}). This crossover is especially prevalent within the size-range we investigate since many of the objects seen impacting the atmosphere by the DFN have $T_{J}$ values around $\sim3$, which is approximately the boundary typically used to separate asteroids and JFCs. 

Within this study, we were particularly interested in assessing the sub-population of Earth-scattered objects that are redirected from asteroid-like orbits ($T_{J}>3$) to JFC-like orbits ($2\geq T_{J} \leq 3$). Fig.~\ref{fig:tj_ba_scatter} shows there are quite a significant fraction of close encounters that undergo this transformation in region A. In total, we calculate the net annual object flux for this population is about $2.5 \times 10^{8}$\,objects per year. This number also takes into account objects coming from JFC-like orbits onto asteroidal ones (region B in Fig.~\ref{fig:tj_ba_scatter}). This flux is net positive onto JFC-like orbits because the most likely objects to have a close encounter with the Earth are objects with orbits most similar to the orbit of the Earth. As shown in \citet{carusi1996close}, since orbits of near-Earth asteroids are more stable than JFCs and have orbits more like the Earth's, they are likely to encounter the Earth more regularly. 

For a more useful interpretation of the same data shown in Fig.~\ref{fig:tj_ba_scatter}, please refer to the Appendix where there are three tables describing the mass and object flux. 

\subsection{Cumulative Size-Frequency Distributions}

From our model, we were able to estimate the cumulative size-frequency distribution (CSD) for the terrestrial close encounter population along with its sub-populations of interest. Annually, we have found that within $1.5$, $0.5$, and $0.1$\,LD the average largest encounter predicted respectively are $>100$\,m, $\sim50$\,m, and $10-20$\,m in diameter. A more thorough analysis concerning the entire DFN's CSD will be addressed in a separate study (Sansom et al. in prep.). For the remainder of this study, we have focused on understanding the redirected smaller sub-population modelled. 

\begin{figure}
    \centering
    \includegraphics[width=\columnwidth]{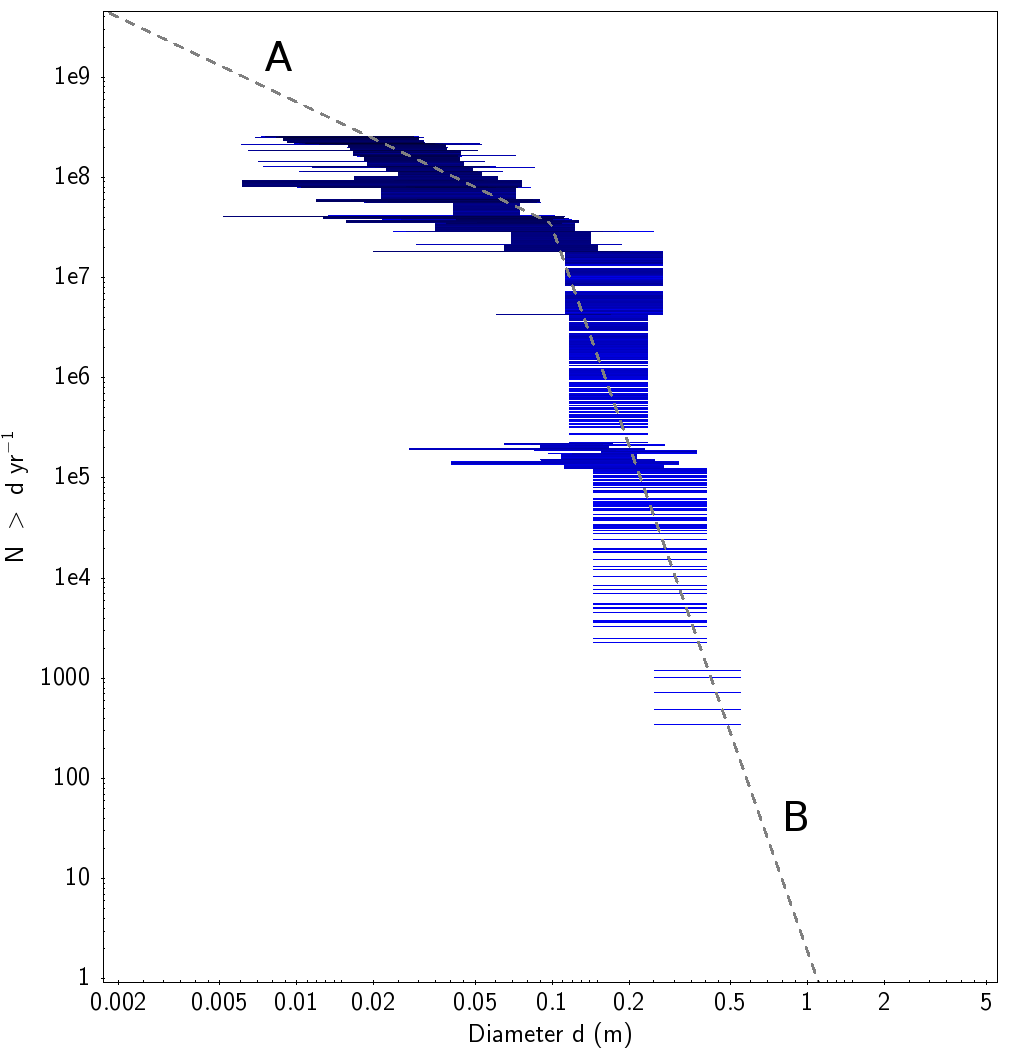}
    \caption{Cumulative size frequency distribution of annual close encounters $<1.5$\,LD redirected from asteroidal to JFC-like orbits. Horizontal lines are indicative of uncertainty in the diameter. This sub-population only represents $\sim0.1\%$ of the close encounters given the maximum MOID. The slopes of branches A and B are $-1.21\pm0.01$ and $-7.22\pm1.30$ respectively. Compared to the entire close encounter flux, the B branch is steeper and the A branch is shallower. There are proportionally more higher mass meteoroids.}
    \label{fig:aster2jfc}
\end{figure}

\begin{figure}
    \centering
    \includegraphics[width=\columnwidth]{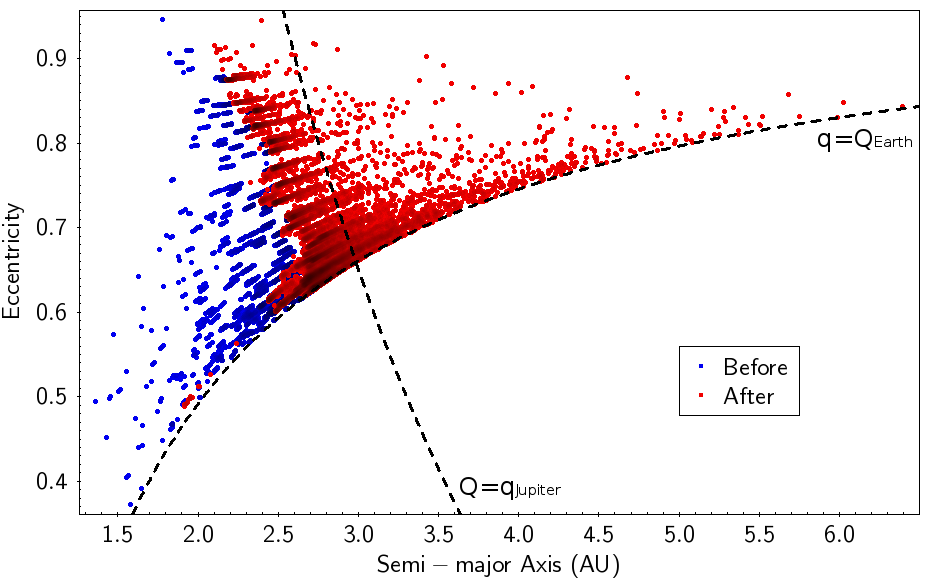}
    \caption{Semi-major axis vs. eccentricity for particles that were transferred from asteroidal like orbits ($T_{J}>3$) to JFC-like orbits ($2<T_{J}<3$). As shown by the particles lying on or past the perihelion of Jupiter, close encounters transfer objects onto orbits indistinguishable at times from native JFCs.}
    \label{fig:ba_2jfc}
\end{figure}

Similar to \citet{bland2006rate}, when characterizing the resulting CSDs from the close encounter model, the range was split into appropriate branches. These branches each have distinct slopes, and these are indicative of some underlying change in the production or physical evolution of objects within that size range. As seen in Fig.~\ref{fig:aster2jfc}, we have split the CSD for objects going from asteroidal to JFC orbits into two branches. These branches, labeled A and B, have slopes of $-1.21\pm0.01$ and $-7.22\pm1.30$, respectively. This trend, a shallower slope and then a sudden increase in slope at $\sim3$\,kg, is very similar to the top of the atmosphere flux found in \citet{bland2006rate} within the given size-range. However, the estimated slopes for the asteroidal to JFC-like flux are different than those in \citet{bland2006rate}, indicating some size-dependence. The slope for branch A ($<0.1$\,m in diameter), is slightly shallower for the scattered asteroidal to JFC population, $-0.410\pm0.001$ vs. $-0.480$ with log(mass) as the x-axis. Whereas branch B ($>0.1$\,m in diameter) is exceptionally steep compared to the same size range in \citet{bland2006rate}, $-2.277\pm0.103$ vs. $-0.926$ with log(mass) as the x-axis. This increase in the discrepancy between branches A and B is indicative that this sub-population (compared to all close encounters) is proportionally more weighted towards larger masses. This size sorting likely is caused by the increase in average mass for meteoroids from asteroidal sources compared to cometary ones. Moreover, these meteoroids are dispersed on a multitude of orbits, some of which may be indistinguishable from actual JFCs (Fig.~\ref{fig:ba_2jfc}) as they are likely to have multiple close encounters with Jupiter in their lifetime \citep{tancredi2014criterion}. 

In the metre size-range, there is no consistent mechanism to produce meteoroids from comets besides catastrophic breakup \citep{jenniskens2004,jenniskens2005meteor}. Furthermore, even if there exist centimetre-metre sized objects from breakup events, the physical lifetimes are predicted to be extremely short \citep{levison1997kuiper,beech2001endurance,boehnhardt2004split}. Whereas, this group of predominantly genetically `asteroidal' material now on JFC-like orbits could survive much longer than any cometary meteoroids. Assuming the physical lifetime of these objects is much longer than the dynamic lifetime, these scattered meteoroids should exist on JFC-like orbits for about $10^{5}-10^{6}$\,years based on previous dynamical models \citep{nugent2012detection,vokrouhlicky2015yarkovsky,granvik2018debiased}. This would result in a steady-state population of scattered asteroidal meteoroids on JFC-like orbits of about $10^{13}-10^{14}$ objects annually with perihelia near the Earth. Considering the uncertainty of the CSD slope of this population, if extrapolated, the most massive steady-state object is predicted to be $10^{0}-10^{2}$\,m in diameter based on our model. 

Multiple studies have also found that there exists high-albedo ($p_{v}>0.1$) asteroids on JFC-like orbits in near-Earth space (perihelion distance, $q<1.3$\,au) with diameters $<3$\,km \citep{kim2014physical,licandro2016size}. These authors argued that these objects could have migrated into this region via non-gravitational effects (such as the Yarkovsky effect) due to their smaller size, higher reflectance, and smaller perihelion distance. These objects can also be transferred to JFC-like orbits directly via MMRs in the outer MB (like the 2:1 and 9:4 MMRs) \citep{fernandez2002there,fernandez2014assessing,fernandez2015jupiter,hsieh2020potential}. While most outer-MB objects are dark with low-albedos ($p_{v}<0.1$), some higher-albedo objects have also been observed to exist \citep{masiero2014main}. It is uncertain whether this population is large/efficient enough to account for the high-albedo objects on JFC-like orbits near the Earth. The high-albedo objects analysed in \citet{kim2014physical} also only seem to be present for bodies with $q<1.3$\,au - an observation not due to detection bias as high-albedo objects with larger perihelia would have been detected. Alternatively, these high-albedo near-Earth objects in JFC-like orbits could result from close encounters with the Earth. In \citet{hsieh2020potential}, they found that close encounters with the terrestrial planets or non-gravitational forces were capable of generating a reasonable number of objects from the Themis family to go onto JFC-like orbits. This would not be entirely unprecedented; encounters with the Earth at the same scales have also been linked with refreshing the surfaces of some asteroids \citep{binzel2010earth}. In \citet{bland2006rate}, they found that the slope of the CSD for impactors of the upper atmosphere decreased significantly as the objects grew to $1.7\times10^{10}$\,kg ($\sim88$\,m assuming a $3500$\,kg\,$m^{-3}$ spheroid). Therefore, if this trend persists within the population scattered onto JFC orbits, it is possible that this population could provide an explanation for the asteroids observed by \citet{kim2014physical}. However, given the uncertainty in slope B (Fig.~\ref{fig:aster2jfc}), this cannot be shown in this study.


\subsection{Impact Frequency}
Finally, we wanted to estimate the impact frequency of the population of objects scattered from asteroid to JFC-like orbits. The impact frequency is crucial as it determines whether this meteoroid population is observable by fireball networks. Thus, it could partially explain the durable JFC fireballs that have been observed. 

We employed the methodology described in \citet{greenberg1982orbital} and \citet{bottke1994velocity}, where the impact and close encounter probabilities were determined for Earth and Jupiter geometrically assuming uniform precession of nodes and apsides. For a concise explanation of this methodology, please refer to Appendix A found in \citet{lefeuvre2008nonuniform}. To estimate the impact frequency, we calculated the 3 Hill radii close encounter frequency with Jupiter for all the test particles and the impact frequency with the Earth. Particles with close encounters with Jupiter were assumed to be removed from the Earth-crossing population during the encounter. In contrast, particles that avoid having close encounters with Jupiter were assumed to have Earth-crossing lifetimes of $10^{5}-10^{6}$ according to published outer-MB NEO dynamical lifetimes \citep{nugent2012detection,vokrouhlicky2015yarkovsky,granvik2018debiased}. 

Upon investigation, we found that only $\sim30\%$ of the asteroidal to JFC scattered population were capable of having close encounters with Jupiter. Thus, despite all the particles being within the range $2<T_{J}<3$, most of the objects likely still evolve on very predictable asteroid-like orbits like described in \citet{tancredi2014criterion}. The $30\%$ that do experience close encounters do so every 10-200\,years, with the median time being 50\,years. These objects will likely evolve in a way indistinguishable from a comet from the scattered-disk, similar to the grazing meteoroid observed by the DFN in 2017 \citep{shober2020did}. However, given the shorter residence times, this unstable portion of the scattered population makes up minor fraction of the steady-state population. 

Given the annual flux onto cometary orbits along with the estimated residence times from encounter and impact frequencies, we find that there are approximately $10^{3}-10^{4}$ impacts annually originating from this population. At a minimum, this means that the DFN should observe (taking into account daylight hours, weather, and station malfunctions) $1-10$ events per year. Based on our model, this minimum value indicates that this population is likely observed regularly by meteor and fireball networks. The test particles with the highest probabilities of impacting the Earth after being scattered were on orbits with low-inclinations ($<5^{\circ}$), lower eccentricity, and perihelia near 1\,au. Therefore, we predict that fireball networks should observe meteoroids that were previously scattered; however, all will be dynamically very consistent with asteroidal debris. 

\subsection{Additional Results}
In addition to flux onto JFC-like orbits, we also found a net-positive annual flux onto Aten-type orbits. For similar reasons to the net flux onto JFC orbits, the model suggests it is slightly more favourable to go from an Apollo-type to an Aten-type orbit than the reverse. In total, we estimate about $10^{7}$ objects are annually transferred to Aten-type orbits via a close encounter, but due to the low number statistics this value is tenuous. Unfortunately, we were unable to further explore this sub-population in detail as the uncertainties for the CSD were considerable. However, it is possible that some of these may eventually evolve onto Atira-type orbits via some combination of planetary perturbations or close encounters. The objects could also evolve periodically between classes if in a Kozai resonance \citep{greenstreet2012orbital}. 

Small debris in the Solar System always eventually impacts a planet, impacts the Sun, or is ejected from the Solar System. Most of the material that is ejected is typically through close encounters with Jupiter, but our model also predicts a modest ejection rate of $10^4$\,objects annually resulting from close encounters with the Earth. This value is very tentative due to the small sample size of such objects in the model input data. 

\begin{figure}
    \centering
    \includegraphics[width=\columnwidth]{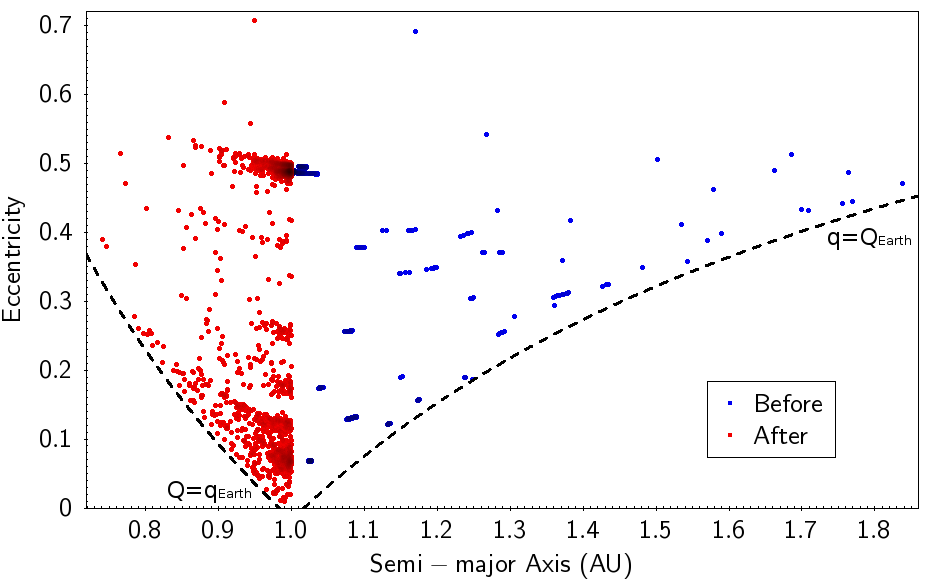}
    \caption{Semi-major axis vs. eccentricity for particles in model that were transferred onto Aten-type orbits as a result of the close encounter with the Earth. Meteoroids with similar orbits to the Earth ($a\sim1-2$\,au), are the most likely to undergo this process.}
    \label{fig:ba_aten}
\end{figure}

\section{Future Work}
We have also studied the dynamical and physical characteristics of $>0.01$\,kg meteoroids from the DFN dataset with JFC-like orbits (Shober et al. in prep.), In this other study, we compared the results from the close encounter model to DFN observations, and tested our hypothesis that asteroidal material is dominant for orbits with $2<T_{J}<3$. In the future, we would also like to check our model by using steady-state NEO models (such as \citet{granvik2018debiased}) to characterise the close encounter flux at the Earth. 

\section{Conclusions}
In this study, we produced a model of close terrestrial encounters within the $0.01-100$\,kg size-range by using DFN data along with global flux studies \citep{bland2006rate}. Close encounters within the model were limited to $<1.5$\,LD, as farther encounters are less likely to have significant orbital alterations. The primary results include: 
\begin{itemize}
    \item there are approximately $1.6 \times 10^{11}$ objects $0.01-100$\,kg that have encounters within $1.5$\,LD of the Earth every year based on extrapolating top of the atmosphere flux rates found in \citet{bland2006rate}.
    \item $2.5\times 10^{8}$\,objects annually are estimated to be transferred from asteroidal orbits ($T_{J}>3$) to JFC-like orbits ($2<T_{J}<3$) ($0.1\%$ of total flux).
    \begin{itemize}
        \item given a dynamic lifetime of $10^{5}-10^{6}$\,years \citep{nugent2012detection,vokrouhlicky2015yarkovsky,granvik2018debiased}, then there exists a steady-state population of $10^{13}-10^{14}$\,objects. 
        \item using the methodologies described by \citet{greenberg1982orbital} and \citet{bottke1994velocity}, we calculated the 3 Hill-radii encounter for Jupiter and impact frequencies for the Earth for all particles transferred from $T_{J}>3$ to $2<T_{J}<3$. Using these values along with the predicted population size from the model, we found that $\sim30\%$ of objects regularly encounter Jupiter and likely evolve chaotically on short-timescales. The remaining $\sim70\%$ of objects avoid close encounters and persist on more stable orbits, evolving in a characteristically asteroidal manner. 
        \item approximately $10^{3}-10^{4}$ impacts occur from this population annually. Thus, the DFN should observe $1-10$ fireballs every year with $2<T_{J}<3$ transferred from the MB. 
        \item extrapolating our CSD slope to larger sizes, the largest steady-state object is $10^{0}-10^{2}$\,m in diameter given the slope uncertainty ($-7.22\pm1.30$).
    \end{itemize}
    \item The model predicts that $\sim10^{7}$ objects are annually transferred onto Aten-type orbits, where some may evolve to Atira-type orbits. 
    \item $\sim10^{4}$ meteoroids are predicted to be directly ejected from the Solar System annually via a close encounter with the Earth, but this value is questionable due to small number statistics. 
\end{itemize}

Observations of fireballs can be used to understand the sources of meteorites, and connect them to the observed asteroidal (or possibly cometary) parent body that they originate from. The speculation over whether or not cometary meteorites exist or can exist has been discussed for decades. However, in this study we have found there should be many genetically asteroidal meteoroids on JFC-like orbits due to close encounters with the Earth that likely impact the Earth regularly. Compounded with material diffusing out from the MB \citep{fernandez2015jupiter,hsieh2020potential}, it may be expected that meteorites from JFC-like orbits ($2<T_{J}<3$) would be quite ordinary.

\section*{Acknowledgements}
We are thankful for the reviewer M. Campbell-Brown and her valuable comments on the original manuscript which improved the paper. The authors would also like to thank P. Brown for his assistance and advice during the early stages of this study. This work was funded by the Australian Research Council as part of the Australian Discovery Project scheme (DP170102529). SSTC authors acknowledge institutional support from Curtin University. This work was supported by resources provided by the Pawsey Supercomputing Centre with funding from the Australian Government and the Government of Western Australia. This research made use of TOPCAT for visualisation and figures \citep{taylor2005topcat}. This research also made use of Astropy, a community-developed core Python package for Astronomy \citep{robitaille2013astropy}. Simulations in this paper made use of the REBOUND code which is freely available at http://github.com/hannorein/rebound \citep{2012AAP_rein}.

\section{Data availability}
The data underlying this article were accessed from Desert Fireball Network (\url{https://dfn.gfo.rocks/}). The derived data generated in this research will be shared on reasonable request to the corresponding author.




\bibliographystyle{mnras}
\bibliography{close_encounter} 




\appendix

\section{Close Encounter Flux Tables}

\begin{landscape}
\begin{table}
\centering
\scriptsize
\caption{Close encounter object flux table. Columns represent the $T_{J}$ value before, and the rows represent the $T_{J}$ value after the close encounter.}
\begin{tabular}{|c|c|c|c|c|c|c|c|c|c|c|c|c|c|c|c|c|c|}
\hline
 & $T_{J}\leq-1$ & $-1-(-0.5)$ & $-0.5-0$ & $0-0.5$ & $0.5-1.0$ & $1.0-1.5$ & $1.5-2.0$ & $2.0-2.5$ & $2.5-3.0$ & $3.0-3.5$ & $3.5-4.0$ & $4.0-4.5$ & $4.5-5.0$ & $5.0-5.5$ & $5.5-6.0$ & $6.0-6.5$ & $T_{J} \geq 6.5$ \\ \hline
$T_{J}\leq-1$ & 1.14e9 & -      & -      & -      & -      & -      & -      & -      & -       & -       & -       & -       & -      & -      & -      & -      & -      \\ \hline 
$-1-(-0.5)$   & 1.28e4 & 2.33e8 & 1.27e4 & -      & -      & -      & -      & -      & -       & -       & -       & -       & -      & -      & -      & -      & -      \\ \hline
$-0.5-0$      & -      & 9.01e6 & 6.22e8 & 3.22e3 & -      & -      & -      & -      & -       & -       & -       & -       & -      & -      & -      & -      & -      \\ \hline
$0-0.5$       & -      & -      & 3.15e3 & 3.13e8 & 1.50e6 & 5.63e2 & -      & -      & -       & -       & -       & -       & -      & -      & -      & -      & -      \\ \hline
$0.5-1.0$     & -      & -      & -      & -      & 1.07e9 & 2.19e5 & 1.54e3 & -      & -       & -       & -       & -       & -      & -      & -      & -      & -      \\ \hline
$1.0-1.5$     & -      & -      & -      & -      & 4.98e3 & 9.91e8 & 1.15e6 & 1.24e3 & 4.67e3  & 2.18e2  & -       & -       & -      & -      & -      & -      & -      \\ \hline
$1.5-2.0$     & -      & -      & -      & -      & -      & 1.06e7 & 2.33e9 & 1.36e6 & 5.53e4  & 2.05e4  & -       & -       & -      & -      & -      & -      & -      \\ \hline
$2.0-2.5$     & -      & -      & -      & -      & -      & 4.74e2 & 1.31e6 & 8.25e9 & 2.20e8  & 2.50e5  & 1.26e4  & 1.35e3  & -      & -      & -      & -      & -      \\ \hline
$2.5-3.0$     & -      & -      & -      & -      & -      & -      & 3.72e3 & 3.54e8 & 3.39e10 & 1.19e9  & 1.66e5  & 9.76e3  & 2.15e3 & -      & -      & -      & -      \\ \hline
$3.0-3.5$     & -      & -      & -      & -      & -      & -      & -      & 2.50e4 & 9.40e8  & 5.11e10 & 8.46e8  & 1.80e5  & 1.80e3 & 7.39e2 & -      & -      & -      \\ \hline
$3.5-4.0$     & -      & -      & -      & -      & -      & -      & -      & 4.90e3 & 2.00e5  & 7.73e8  & 1.99e10 & 4.90e8  & 9.20e4 & 6.80e3 & 8.74e2 & -      & -      \\ \hline
$4.0-4.5$     & -      & -      & -      & -      & -      & -      & -      & -      & 3.35e4  & 5.17e5  & 1.98e8  & 1.31e10 & 2.29e8 & 1.26e5 & 2.98e3 & 3.03e3 & -      \\ \hline
$4.5-5.0$     & -      & -      & -      & -      & -      & -      & -      & -      & 3.92e3  & 5.52e4  & 3.65e5  & 1.56e8  & 6.96e9 & 8.25e7 & 2.66e5 & 5.00e3 & -      \\ \hline
$5.0-5.5$     & -      & -      & -      & -      & -      & -      & -      & -      & -       & 1.48e4  & 4.11e4  & 2.30e5  & 3.98e7 & 3.34e9 & 1.36e8 & 5.38e4 & -      \\ \hline
$5.5-6.0$     & -      & -      & -      & -      & -      & -      & -      & -      & -       & -       & 7.65e3  & 2.36e4  & 1.00e5 & 5.24e7 & 2.55e9 & 4.26e7 & -      \\ \hline
$6.0-6.5$     & -      & -      & -      & -      & -      & -      & -      & -      & -       & -       & 7.22e2  & 9.57e3  & 2.27e4 & 1.16e5 & 6.70e7 & 1.29e9 & -      \\ \hline
$T_{J} \geq 6.5$ &  -  & -      & -      & -      & -      & -      & -      & -      & -       & -       & -       & -       & -      & -      & -      & -      & 3.22e9 \\ \hline

\end{tabular}
\end{table}
\end{landscape}

\begin{landscape}
\begin{table}
\centering
\scriptsize
\caption{Close encounter mass flux table (kg). Columns represent the $T_{J}$ value before, and the rows represent the $T_{J}$ value after the close encounter.}
\begin{tabular}{|c|c|c|c|c|c|c|c|c|c|c|c|c|c|c|c|c|c|}
\hline
 & $T_{J}\leq-1$ & $-1-(-0.5)$ & $-0.5-0$ & $0-0.5$ & $0.5-1.0$ & $1.0-1.5$ & $1.5-2.0$ & $2.0-2.5$ & $2.5-3.0$ & $3.0-3.5$ & $3.5-4.0$ & $4.0-4.5$ & $4.5-5.0$ & $5.0-5.5$ & $5.5-6.0$ & $6.0-6.5$ & $T_{J} \geq 6.5$ \\ \hline
$T_{J}\leq-1$ & 8.47e7 & -      & -      & -      & -      & -      & -      & -       & -       & -       & -       & -       & -       & -      & -      & -      & -      \\ \hline 
$-1-(-0.5)$   & 2.54e3 & 4.89e7 & 2.14e3 & -      & -      & -      & -      & -       & -       & -       & -       & -       & -       & -      & -      & -      & -      \\ \hline
$-0.5-0$      & -      & 1.89e6 & 5.92e7 & 5.91e1 & -      & -      & -      & -       & -       & -       & -       & -       & -       & -      & -      & -      & -      \\ \hline
$0-0.5$       & -      & -      & 5.45e1 & 5.74e6 & 2.46e5 & 1.30e1 & -      & -       & -       & -       & -       & -       & -       & -      & -      & -      & -      \\ \hline
$0.5-1.0$     & -      & -      & -      & -      & 1.48e8 & 5.44e3 & 4.40e2 & -       & -       & -       & -       & -       & -       & -      & -      & -      & -      \\ \hline
$1.0-1.5$     & -      & -      & -      & -      & 9.35e2 & 8.52e8 & 3.46e5 & 2.86e3  & 1.06e3  & 1.76e1  & -       & -       & -       & -      & -      & -      & -      \\ \hline
$1.5-2.0$     & -      & -      & -      & -      & -      & 2.38e7 & 1.28e9 & 1.19e6  & 7.01e4  & 1.08e4  & -       & -       & -       & -      & -      & -      & -      \\ \hline
$2.0-2.5$     & -      & -      & -      & -      & -      & 8.94e2 & 1.80e5 & 2.24e10 & 8.73e7  & 3.73e5  & 1.40e4  & 4.29e2  & -       & -      & -      & -      & -      \\ \hline
$2.5-3.0$     & -      & -      & -      & -      & -      & -      & 3.38e3 & 1.65e8  & 5.42e10 & 4.93e8  & 1.01e5  & 1.24e4  & 4.67e2  & -      & -      & -      & -      \\ \hline
$3.0-3.5$     & -      & -      & -      & -      & -      & -      & -      & 4.99e4  & 3.76e8  & 7.75e10 & 3.93e8  & 3.64e5  & 1.44e3  & 6.73e2 & -      & -      & -      \\ \hline
$3.5-4.0$     & -      & -      & -      & -      & -      & -      & -      & 2.29e2  & 1.51e5  & 2.61e8  & 2.03e10 & 1.06e9  & 5.60e5  & 2.15e3 & 2.62e1 & -      & -      \\ \hline
$4.0-4.5$     & -      & -      & -      & -      & -      & -      & -      & -       & 2.54e4  & 3.42e5  & 2.31e8  & 2.13e10 & 8.51e7  & 4.15e5 & 1.91e3 & 1.57e3 & -      \\ \hline
$4.5-5.0$     & -      & -      & -      & -      & -      & -      & -      & -       & 1.57e3  & 4.34e4  & 2.48e5  & 7.72e7  & 1.87e10 & 4.19e8 & 2.88e5 & 9.55e2 & -      \\ \hline
$5.0-5.5$     & -      & -      & -      & -      & -      & -      & -      & -       & -       & 5.61e3  & 1.61e4  & 2.25e5  & 8.22e7  & 8.95e9 & 1.99e7 & 1.22e5 & -      \\ \hline
$5.5-6.0$     & -      & -      & -      & -      & -      & -      & -      & -       & -       & -       & 2.55e3  & 1.67e4  & 4.43e5  & 9.22e6 & 3.84e9 & 4.48e7 & -      \\ \hline
$6.0-6.5$     & -      & -      & -      & -      & -      & -      & -      & -       & -       & -       & 4.77e2  & 5.54e3  & 3.35e4  & 2.02e5 & 3.70e8 & 4.05e9 & -      \\ \hline
$T_{J} \geq 6.5$ & -   & -      & -      & -      & -      & -      & -      & -       & -       & -       & -       & -       & -       & -      & -      & -      & 6.69e9 \\ \hline
\end{tabular}
\end{table}
\end{landscape}

\begin{landscape}
\begin{table}
\centering
\scriptsize
\caption{Close encounter mass error flux table (kg). Columns represent the $T_{J}$ value before, and the rows represent the $T_{J}$ value after the close encounter.}
\begin{tabular}{|c|c|c|c|c|c|c|c|c|c|c|c|c|c|c|c|c|c|}
\hline
 & $T_{J}\leq-1$ & $-1-(-0.5)$ & $-0.5-0$ & $0-0.5$ & $0.5-1.0$ & $1.0-1.5$ & $1.5-2.0$ & $2.0-2.5$ & $2.5-3.0$ & $3.0-3.5$ & $3.5-4.0$ & $4.0-4.5$ & $4.5-5.0$ & $5.0-5.5$ & $5.5-6.0$ & $6.0-6.5$ & $T_{J} \geq 6.5$ \\ \hline
$T_{J}\leq-1$ & 7.57e6 & -      & -      & -      & -      & -      & -      & -      & -       & -       & -      & -      & -      & -      & -      & -      & -      \\ \hline 
$-1-(-0.5)$   & 1.14e2 & 2.84e7 & 1.93e4 & -      & -      & -      & -      & -      & -       & -       & -      & -      & -      & -      & -      & -      & -      \\ \hline
$-0.5-0$      & -      & 1.10e6 & 4.37e8 & 1.44e2 & -      & -      & -      & -      & -       & -       & -      & -      & -      & -      & -      & -      & -      \\ \hline
$0-0.5$       & -      & -      & 2.38e1 & 1.40e7 & 3.69e4 & 2.77e1 & -      & -      & -       & -       & -      & -      & -      & -      & -      & -      & -      \\ \hline
$0.5-1.0$     & -      & -      & -      & -      & 6.88e7 & 1.20e4 & 1.89e2 & -      & -       & -       & -      & -      & -      & -      & -      & -      & -      \\ \hline
$1.0-1.5$     & -      & -      & -      & -      & 1.29e2 & 5.68e8 & 1.59e5 & 3.46e2 & 3.93e3  & 1.35e0  & -      & -      & -      & -      & -      & -      & -      \\ \hline
$1.5-2.0$     & -      & -      & -      & -      & -      & 1.50e7 & 1.98e8 & 1.62e5 & 1.51e4  & 1.46e3  & -      & -      & -      & -      & -      & -      & -      \\ \hline
$2.0-2.5$     & -      & -      & -      & -      & -      & 5.89e2 & 2.61e4 & 3.52e9 & 7.03e6  & 3.26e4  & 9.60e2 & 1.29e1 & -      & -      & -      & -      & -      \\ \hline
$2.5-3.0$     & -      & -      & -      & -      & -      & -      & 5.21e3 & 1.54e7 & 1.10e10 & 5.50e8  & 1.95e4 & 1.67e3 & 2.19e1 & -      & -      & -      & -      \\ \hline
$3.0-3.5$     & -      & -      & -      & -      & -      & -      & -      & 8.82e3 & 1.28e8  & 1.88e10 & 1.04e8 & 6.87e4 & 9.41e1 & 2.63e2 & -      & -      & -      \\ \hline
$3.5-4.0$     & -      & -      & -      & -      & -      & -      & -      & 1.61e1 & 1.05e5  & 3.95e7  & 3.89e9 & 1.71e8 & 2.43e5 & 4.51e2 & 2.06e0 & -      & -      \\ \hline
$4.0-4.5$     & -      & -      & -      & -      & -      & -      & -      & -      & 3.33e3  & 4.24e4  & 6.09e7 & 3.04e9 & 1.22e7 & 1.81e5 & 5.94e2 & 6.87e2 & -      \\ \hline
$4.5-5.0$     & -      & -      & -      & -      & -      & -      & -      & -      & 6.46e2  & 7.72e3  & 9.29e4 & 1.07e7 & 6.95e9 & 1.81e8 & 1.23e5 & 7.96e1 & -      \\ \hline
$5.0-5.5$     & -      & -      & -      & -      & -      & -      & -      & -      & -       & 7.41e2  & 5.77e3 & 4.36e4 & 1.85e7 & 1.42e9 & 6.91e6 & 6.64e4 & -      \\ \hline
$5.5-6.0$     & -      & -      & -      & -      & -      & -      & -      & -      & -       & -       & 4.74e2 & 3.42e3 & 1.63e5 & 2.99e6 & 1.12e9 & 2.22e7 & -      \\ \hline
$6.0-6.5$     & -      & -      & -      & -      & -      & -      & -      & -      & -       & -       & 3.97e1 & 1.44e3 & 1.50e4 & 8.06e4 & 7.13e7 & 1.68e9 & -      \\ \hline
$T_{J} \geq 6.5$ & -   & -      & -      & -      & -      & -      & -      & -      & -       & -       & -      & -      & -      & -      & -      & -      & 9.24e8 \\ \hline
\end{tabular}
\end{table}
\end{landscape}


\bsp	
\label{lastpage}
\end{document}